\documentclass[pre,floatfix,superscriptaddress,twocolumn,showpacs]{revtex4-1}
\usepackage{graphicx}
\usepackage{amsmath,amssymb,amsfonts}
\usepackage{mathtools}
\usepackage{bm}
\usepackage{color}
\usepackage[colorinlistoftodos]{todonotes}

\begin{document}

\title{Self-Assembly of Active Bifunctional Patchy Particles}
\author{Caterina Landi}
\affiliation{Departamento de Estructura de la Materia, Física Térmica y Electrónica, Universidad Complutense de Madrid, 28040 Madrid, Spain}
\author{John Russo}
\affiliation{Department of Physics, Sapienza Università di Roma, Piazzale Aldo Moro 5, 00185 Rome Italy.}
\author{Francesco Sciortino}
\affiliation{Department of Physics, Sapienza Università di Roma, Piazzale Aldo Moro 5, 00185 Rome Italy.}
\author{Chantal Valeriani}
\affiliation{Departamento de Estructura de la Materia, Física Térmica y Electrónica, Universidad Complutense de Madrid, 28040 Madrid, Spain}

\begin{abstract}
In this work, with the intent of exploring the out-of-equilibrium polymerization of active patchy particles in linear chains, we study a suspension of active bifunctional Brownian particles (ABBPs). At all studied temperatures and densities, ABBPs self-assemble in aggregating chains, as opposed to the uniformly space-distributed chains observed in the corresponding passive systems.
The main effect of activity, other than inducing chain aggregation, is to reduce the chain length and favour alignment of the propulsion vectors in the bonding process. At low activities, attraction dominates over activity in the bonding process, leading self-assembly to occur randomly regardless of the particle orientations.
Interestingly, we find that at the lowest  temperature, as density increases, chains aggregate forming a novel state: MISP, i.e., Motility-Induced Spirals, where spirals are characterised by a finite angular velocity. 
On the contrary, at the highest  temperature, density and activity, chains aggregate forming  a different novel state (a spinning crystalline cluster) characterised by 
a compact and hexagonally ordered structure, both translating and rotating. The rotation arises from an effective torque generated by the presence of competing domains where particles self-propel in the same direction.
\end{abstract}

\maketitle

\section{Introduction}

In the last decades, significant progress in the comprehension of the structural and dynamical properties of liquids has been made through the investigation of colloidal particles interacting via spherically symmetric or (more realistic) anisotropic forces ~\cite{hansen2013theory}. 
A practical model proposed to study anisotropic interactions between colloidal particles is the so-called ``patchy particle'' model, consisting of hard-spheres whose surface is decorated with a finite number of short-range attractive sites ~\cite{Bianchi2011, Rovigatti2018}.
Patchy particles have allowed elucidating the behavior of network-forming materials ~\cite{sciortino2008gel, russo2022physics}, such as water ~\cite{Nezbeda1989} or silica ~\cite{Ford2004}, finite aggregates, such as surfactant micelles ~\cite{Kraft2012}, or more complex structures, such as proteins ~\cite{Sear1999,liu2007vapor}. 
Patchy particles have also represented a novel class of building blocks for constructing precise structures, where the arrangements and the selectivity of the sites dictate the overall structure of the assemblies ~\cite{Noya2014, Tracey2021, Noya2021,liu2024inverse}. 
Indeed, the bottom-up approach of patchy colloidal self-assembly has proven to be pivotal for technological advancements across diverse fields, including materials science ~\cite{Material_science}, pharmaceutical industry ~\cite{Pharmaceutical_science}, electronics ~\cite{Electronics}, nanotechnology ~\cite{boal2000}, and even food technology ~\cite{Food_science}.

Optimization of colloidal self-assembly, inspired by biological matter, has been further achieved by introducing activity ~\cite{ActiveMatterReview} on simple colloids, with possible applications ranging from targeted drug delivery ~\cite{Drug_delivery} to autonomous depollution of contaminated water and soils ~\cite{Pollutants}.

The majority of the published work on active colloidal matter has focused on suspensions of active particles interacting via an isotropic potential (attractive ~\cite{Mognetti2013} or repulsive ~\cite{Stenhammar2015}). Only more recently, the field of active matter has branched out to explore the interplay between activity and anisotropic interactions, with the goal of developing a systematic understanding of how active forces can be exploited together with anisotropic forces to design assemblies with desired structural and functional features ~\cite{Alarcon2019, Mallory2019, Mallory2017}.

\textcolor{black}{
Active colloids have shown to aggregate into functional structures not detected in equilibrium systems. When active particles are spherical and repulsive, they undergo a Motility Induced Phase Separation ~\cite{Stenhammar2015, cates2015}, whereas when spherical and attractive they form living clusters ~\cite{Mognetti2013, Redner2013}. Interestingly, when active particles are elongated, they aggregate into functional transient clusters capable of rotating, such as those reported in ~\cite{Suma2014, Schwarz-Linek2012}. Thus, in order to detect a spinning state in a self-assembled suspension of active particles, one needs two main ingredients: particles’ activity and shape anisotropy.  On the other side, one could consider spinning of an already formed structure. Spinning has also been observed in a passive gear embedded in an active bath of elongated particles, such as a bacterial suspension ~\cite{sokolov2010, 2010bacterial}. When dealing with  active polymers, a spinning spiral state appears whenever the propulsion force along the polymer backbone is tightly parallel to the local tangent ~\cite{janzen2024, Zhu_2024}. A  recent work has revealed a spinning state in a suspension of active particles whose shape is more complex than spherical and is characterized by an attractive patch on their surface ~\cite{Wang2020}.}

Full control of the complex dynamics of active patchy colloids remains yet challenging. 
So far, research has focused on tuning the shape, size, and composition of the patches in order to control  autonomous locomotion and spontaneous assembly ~\cite{Wang2019, Wang2020, Mallory2017}. Specific interactions can be obtained by implementing lock and key groups on the particles’ surfaces, such as DNA oligonucleotides, protein cross-linkers or antibody-antigen binding pairs ~\cite{Wang2012}. Due to their ability to self-assemble into chains, sheets, rings, icosahedra, tetrahedra, etc., patchy colloids provide access to a broad range of active colloidal materials ~\cite{Zhang2004}.

In this article, we explore the effects of activity on a system of active patchy particles forming linear chains ~\cite{Sciortino2007}. In section~\ref{sims}, we report the numerical details of the system under study: a two-dimensional suspension of active Brownian repulsive particles whose surface is decorated with two diametrically-opposed attractive sites that interact via a short-range attractive potential. 
In section~\ref{results}, we report the results, focusing on the structural features and on two observed novel states: active spirals and spinning crystalline clusters ~\cite{holl2024}.

\section{Simulation Details and Analysis Tools}
\label{sims}

\begin{figure}[h!]
 \centering
 \includegraphics[height=3.8cm]{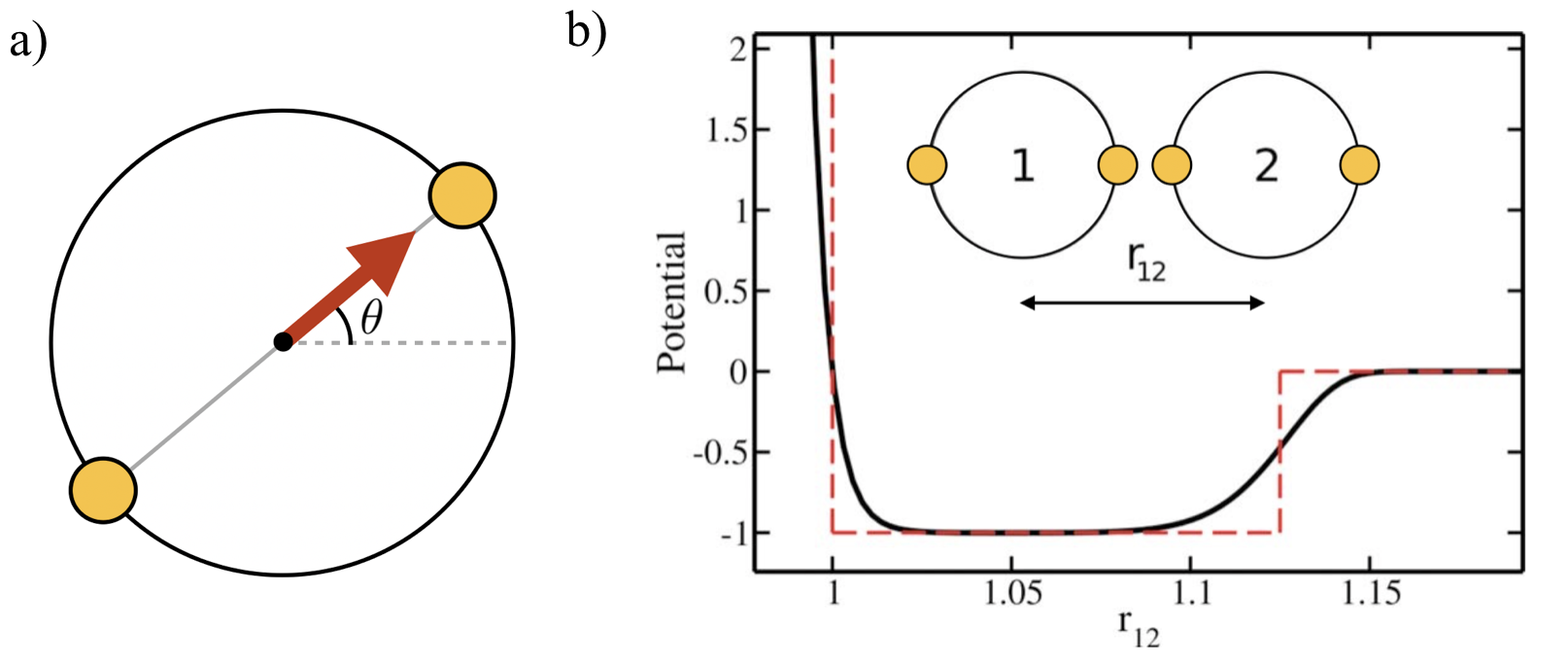}
 \caption{
 a) Pictorial representation of ABBPs: hard-disks featuring two identical and diametrically-opposed attractive sites (small golden disks) and self-propulsion (red arrow) along the segment connecting the two sites.
 b) Interaction potential between two ABBPs in the most favorable bonding configuration, i.e., when two sites are facing each other (see inset).
 The dashed red line represent the hard-core plus square-well potential used as reference for the choice of the interaction potential. 
 }
\label{fgr:ABBP}
\end{figure}

We simulate a two-dimensional system of active bifunctional Brownian particles (ABBPs) in a square box with periodic boundary conditions. Particles are modeled as hard-disks of diameter $\sigma$, featuring two identical and diametrically-opposed attractive sites, and self-propelling in the direction of the vector connecting the two sites (see Fig. \ref{fgr:ABBP}a).

The two-body interaction potential between particles $i$ and $j$ is given by:
\begin{equation}
    V(i,j) = V_{CM}(i,j) + V_S(i,j)
    \label{interaction_potential}
\end{equation}
where $V_{CM}(i,j)$ represents the hard-core interaction between the centers of mass and $V_{S}(i,j)$ the directional attractive interaction between the sites. Specifically, as in Ref. ~\cite{Russo2009}, we choose:
\begin{align}
    & V_{CM}(i,j) = \left(\frac{\sigma}{r_{ij}}\right)^m \\
    & V_S(i,j) = -\sum_{a=1}^2 \sum_{b=1}^2 \epsilon\: \exp \left[\, -\frac{1}{2}\left(\frac{r_{ij}^{\:ab}}{\alpha}\right)^n\;  \right]
\end{align}
where $r_{ij}$ is the distance between the centers of mass of the two particles and $r_{ij}^{\:ab}$ is the distance between sites $a$ and $b$ located on particles $i$ and $j$ respectively. The selected interaction potential incorporates the following assumptions:
1) particles are hard ($m=200$); 
2) the site-site $V_S$ interaction resembles a square-well ($n=10$);
3) the single bond per site condition is fulfilled ($\alpha=0.12$);
4) the potential depth $u_0$ is set to $1$ ($\epsilon=1.001$).
\textcolor{black}{The choice of $\alpha =0.12$ rises from purely geometric considerations. Indeed, geometric considerations for a three touching spheres configuration show that the choice of a well-width $~0.119\sigma$ guarantees that each site is engaged at most in one bond ~\cite{Sciortino2007}.}
Fig. \ref{fgr:ABBP}b depicts the shape of the interaction potential (black line) when two particles are in the most favorable bonding configuration.

Each particle is characterized by the position vector of its center of mass $\mathbf{r}=(x,y,0)$ and the orientation angle $\theta$ representing the direction of the vector connecting the two sites with respect to the $x$-axis. The orientation vector $\eta=(\cos\theta,\sin\theta,0)$ is applied at each particle's center of mass and is restricted to rotate in the two-dimensional plane of the system. 

While self-propelling in the same direction of the orientation vector with a constant speed $v$, each particle undergoes Brownian motion, in both position and orientation, at a constant temperature $T$. Thus, for a particle $i$, the translational and rotational equations of motion, read as:

\begin{align}
    & \dot{\mathbf{r}}_i(t)= \frac{D_T}{K_BT} \,\mathbf{F}_i\left(\{\mathbf{r}_{ij},\boldsymbol{\eta}_i,\boldsymbol{\eta}_j\}\right) + v\mathbf{\boldsymbol{\eta}}_i(t) + \sqrt{2D_T}\mathbf{\boldsymbol{\xi}}_T(t) \\
    & \dot{\boldsymbol{\eta}}_i(t)= \frac{D_R}{K_BT}
    \,\mathbf{T}_i\left(\{\mathbf{r}_{ij},\boldsymbol{\eta}_i,\boldsymbol{\eta}_j\}\right) + \sqrt{2D_R}\boldsymbol{\xi}_R(t) \times \boldsymbol{\eta}(t)
\end{align}
The diffusion coefficients $D_T$ and $D_R$ relate to each other via  $D_R=3D_T/\sigma^2$. In both the translational and rotational equations, the Gaussian white-noise terms are characterized by $<\xi(t)>=0$ and $<\xi(t)\xi(t')>=\delta(t-t')$. The total force $\mathbf{F}_i$ and torque $\mathbf{T}_i$ acting on each particle are respectively given by:
\begin{align}
    & \mathbf{F}_i\left(\{\mathbf{r}_{ij},\boldsymbol{\eta}_i,\boldsymbol{\eta}_j\}\right) =
    \sum_{j\neq i} \mathbf{F}_{ij}(r_{ij},\theta_i,\theta_j)=
    -\sum_{j\neq i} \boldsymbol{\nabla}_{\mathbf{r}_{ij}}
     \, V_{ij}(r_{ij},\theta_i,\theta_j) \\
    & \mathbf{T}_i \left(\{\mathbf{r}_{ij},\boldsymbol{\eta}_i,\boldsymbol{\eta}_j\}\right) =
    \sum_{j\neq i} \mathbf{T}_{ij}(r_{ij},\theta_i,\theta_j) = \sum_{j\neq i} \boldsymbol{\eta}_i\times\frac{\partial \,V_{ij}(r_{ij},\theta_i,\theta_j)}{\partial\,\boldsymbol{\eta}_i}
\end{align}
where $\mathbf{F}_{ij}$ and $\mathbf{T}_{ij}$ are respectively the force and the torque  between particles $i$ and $j$ interacting via the potential $V_{ij}=V(i,j)$ described in Eq. \ref{interaction_potential}. The potential only depends on the distance between centers of mass $r_{ij}$ and the orientations of both particles $\theta_i$ and $\theta_j$.

In this article, all results are reported in reduced units. The unit length  is $\sigma$ (one particle's diameter, which is set to $1$) and the energy unit is $u_0$ (the potential depth, which is also set to $1$). With $k_B=1$, temperature is measured in units of energy. Time is in units of $\sigma^2/D_T$. All simulations are run for at least $10^8$ steps with an integration time step of $10^{-6}$ units.

We set the number of particles to $N=5000$ and simulate the system at four different number densities $\rho=N/A$ (with $A$ the total area): $\rho=0.1,\ 0.2,\ 0.3,$ and $0.4$. Activity is quantified by means of the Péclet number, defined as in Ref. ~\cite{MartinRoca2021}:
\begin{equation}
Pe = \frac{3 v \tau_R}{\sigma}
\end{equation}
being $\tau_R = 1/D_R$ the reorientation time. We fix the value of the rotational diffusion ($D_R=3$) and vary the one of the propulsion speed. Specifically, the Péclet number varies among the following values: $Pe=0,\ 1.66,\ 3.33,\ 5,\ 10,$ and $20$ (corresponding to speeds $v=0,\ 1.66,\ 3.33,\ 5,\ 10,$ and $20$ respectively). Simulations in the passive regime ($Pe = 0$) are performed as reference. We choose to study the behaviour of the system at two temperatures: a lower one $T=0.07$ and a higher one $T=0.1$.

In order to study the assembly features of the suspension, we evaluate the chain length distribution $\rho_{ch}$ and the cluster size distribution $\rho_{cl}$ according to:
\begin{align}
    \rho_{ch}(l)=\ \left< \frac{N_{l}}{\ \sum_{l} N_{l}} \right> 
    \label{chain_distirbution}\\
    \rho_{cl}(s)=\ \left< \frac{N_{s}}{\ \sum_{s} N_{s}} \right>
    \label{cluster_distirbution}
\end{align}
where $N_{l}$ is the number of chains of length $l$, $N_{s}$ is the number of clusters of size $s$, $\sum_{l}$ and $\sum_{s}$ run respectively over all chain lengths and all cluster sizes and $\left< \ ...\ \right>$ averages over steady state configurations. \textcolor{black}{The maximum values of $l$ and $s$ are fixed by the largest chain and largest cluster found in the entire simulation.}
On the one hand, the chain length distribution relies on an energetic criterion: two particles form a chain bond when their interaction energy is lower than $-0.3$ units. On the other hand, the cluster size distribution relies on a geometric criterion: two particles belong to   the same cluster when the distance between their centers of mass is smaller than $1.2$ units. 

To describe the structural properties of the suspension, we compute the system structure factor:
\begin{equation}
\label{structt}
    S(q) = \ \left< \frac{1}{N}\sum_{m=1}^{N}\sum_{n=1}^{N}e^{-i\textbf{q}\cdot(\textbf{r}_m-\textbf{r}_n)} \right>
\end{equation}
where $\textbf{q}$ is the exchanged wave vector, $\textbf{r}_m$ is the coordinate of particle $m$ and $\left< \ ... \ \right>$ averages over steady state configurations.

\textcolor{black}{
In order to be able to extract more detailed conclusions, we have decided to compare the $S(q)$ calculated from the simulations (Eq. \ref{structt}) with a theoretical $S(q)$ representative of an ideal gas of polydisperse straight chains (Eq. \ref{structt11}). 
In the ideal gas limit, correlations between different chains can be neglected, and the structure factor of the system should be provided by the structure of a single chain, weighted by the appropriate chain length distribution:
\begin{equation}
\label{structt11}
    S(q)=\frac{\sum_l\rho_l l S_l(q)}{\sum_l\rho_l l}
\end{equation}
where $S_l(q)$ is the structure factor of a chain of length l:
\begin{equation}
S_l(q)=\left<\frac{1}{l}\sum_{m=1}^l\sum_{n=1}^le^{-i\mathbf{q}\cdot(\mathbf{r}_m-\mathbf{r}_n)}\right>
\end{equation}}\textcolor{black}{which, under the approximation that chains are straight, and averaging over all possibles orientations of a chain, becomes:
\begin{align}
    S_l(q) & = \frac{1}{l} \left[\,l + \sum_{m=1}^{l-1}  (l-m)\left< e^{-iq\sigma m\cos\theta} + e^{iq\sigma m\cos\theta} \right> \,\right] \\
    &= \frac{1}{l} \left[\,l +  \sum_{m=1}^{l-1} 2 (l-m) \left< \cos(q\sigma m \cos\theta) \right> \,\right]  \\
    &= 1+\frac{2}{l}\left[\sum_{m=1}^{l-1}\left(l-m\right)J_0(q \sigma m)\right]
\end{align}
where $J_0(x) = \frac{1}{2\pi}\int_0^{2\pi}\cos(x\cos\theta)d\theta$ represents the Bessel function of order zero.}
\textcolor{black}{Any deviation we observe with respect to the theoretical predictions is due to correlation between chains or bending of the chains.}

To analyze the bonding dynamics, we compare the orientation of the first and the second particle of each chain and assign $+1$ if the orientations are the same and $-1$ if they are not (i.e., we measure the scalar product between the two propulsion vectors and assign $+1$ if the product is greater than $0$ or $-1$ if is less than $0$).
We repeat the same procedure for each  pair of particles in the chain and then sum all values. Hence, for each chain $i$ of length $l$, we obtain a value $B_{i}$ ranging between $+(l-1)$ and $-(l-1)$.  The upper limit $+(l-1)$ represents the  case in which all particles are assembled with the same orientation and the lower limit $-(l-1)$ the case with alternating ones. 
We evaluate the average over all $N_l$ chains of length $l$ ($E_B(l)$) and the variance ($Var_B(l)$) as:
\begin{align}
    & E_B(l)= \frac{\sum_{i=1}^{N_l}B_i}{N_l}
    \label{media} \\
    & Var_B(l)=\frac{\sum_{i=1}^{N_l}B_i^2}{N_l} - E_B^2(l)
    \label{varianza}
\end{align}
\textcolor{black}{This method is intended to determine whether two particles prefer to assemble with propulsion vectors aligned in the same direction, in opposite directions, or randomly, rather than evaluating the chain propulsion. Given the flexibility of the chains, simply summing all propulsion vectors in one direction and subtracting the ones in the opposite direction would not adequately evaluate the chain propulsion.}

When the formed structures are more compact, we evaluate the hexagonal order parameter ~\cite{NelsonHalperin1979}, whose expression, for each particle $m$, is given by:
\begin{equation}
    \psi_m = \frac{1}{k} \sum_{n=1}^{k} e^{\:i\:\!k\theta_{mn}}
\end{equation}
where the sum runs over the $k=6$ nearest neighbors and $\theta_{mn}$ is the angle formed by the vector $\mathbf{r}_{mn}$ and the $x$-axis. \textcolor{black}{In particular, we are interested in:
$\psi = \left< \sum_{m=1}^{N}\psi_m / N\right>$, where $\left< \ ... \ \right>$ average over steady-state configurations.}

In the context of the crystalline structure, predominantly characterized by straight chains, we evaluate the chain propulsion as follows. 
For each chain, we compare the orientation of the first particle with the orientation of each other particle and assign $+1$ if they are the same (scalar product greater than $0$) or $-1$ if they are not (scalar product less than $0$). \textcolor{black}{
Then, chain propulsion is obtained by summing 1 (value for first particle) to all values, once computed its absolute value. We normalize the chain propulsion by dividing it by the chain length, with 0 indicating no propulsion and 1 indicating maximum possible propulsion. Intermediate values provide insight into the chain's propulsion relative to its maximum possible value.
}

\textcolor{black}{
To demonstrate that a phase separation takes place, we assess the local density distribution by applying a Voronoi tessellation to the system ~\cite{Voronoi1908}. Each cell in the Voronoi tessellation corresponds to the area of a particle identified by all points that are closer to that particle than to any other. The reciprocal areas of these Voronoi cells can be interpreted as local densities.
}

\textcolor{black}{
As far as we are aware, one can characterise a spiral-like structure quantifying the number of turns of the chain, by computing either the turning number ~\cite{janzen2024} or the spiral number ~\cite{turning_numer1}. 
The two quantities, although defined in a slightly different way, identically give the same information.}

\textcolor{black}{
The turning number is computed\cite{janzen2024}, for each chain $i$ with length $l$, as:
\begin{align}
    & \chi_i = \frac{1}{2\pi}\sum_{j=1}^{l-1}(\beta_{j+1}-\beta_j) 
\end{align}
where $\beta_j$ is defined by $\hat{t}_j=(\cos\beta_j, \sin\beta_j)$, 
which represent the bond unit vector $\hat{t}_j=(\mathbf{r_{j+1}}-\mathbf{r_j})/|\mathbf{r_{j+1}}-\mathbf{r_j}|$.
Thus, $(\beta_{j+1}-\beta_j)$ gives the angle increment between two consecutive bonds.}
\textcolor{black}{
The turning number defines the transition from an elongated to a spiral state, by quantifying the number of turns of the chain between its two ends: $\chi_i=0$ (no turns), $\chi_i=\pm1$ (one turn), $\chi_i=\pm2$ (two turns), and so on. In particular, we are interested in the average turning number, defined as: $\chi = < \sum_{i=1}^{\bar{N}}|\chi_i| / \bar{N} >$, where $\bar{N}$ is the total number of chains and $\left< \ ... \ \right>$ average over steady state configurations.
}

\textcolor{black}{
The spiral number ~\cite{turning_numer1}, for each chain $i$ with length $l$, is defined as: 
\begin{align}
s_i = \frac{\alpha_l - \alpha_1}{2\pi}
\end{align}
where $\alpha_{l}$ is the bond orientation of the last monomer and $\alpha_1$ of the first one. 
To note that, in this case, the bond orientation $\alpha$ takes into account all full rotations. Therefore, $\alpha$ can be larger than $2\pi$.
The spiral number defines the transition from an elongated to a spiral state, by quantifying the number of turns of the chain between its two ends: $s_i=0$ (no turns), $s_i=\pm1$ (one turn), $s_i=\pm2$ (two turns), and so on.
}
\textcolor{black}{
Thus, the definition of the turning number is equivalent to that of the spiral number (see Supplementary Material).}

\textcolor{black}{
To conclude, we underline that one could also use the end-to-end distance of the chain to characterise the spiral state.}
\textcolor{black}{
Our choice is to focus on the turning number in the main text as a way to characterise the system in a spiral state. Even though results on the spiral number will be reported in the Supplementary Material. 
}

\section{Results }
\label{results}

\begin{figure}[h!]
 \centering
 \includegraphics[height=5.7cm]{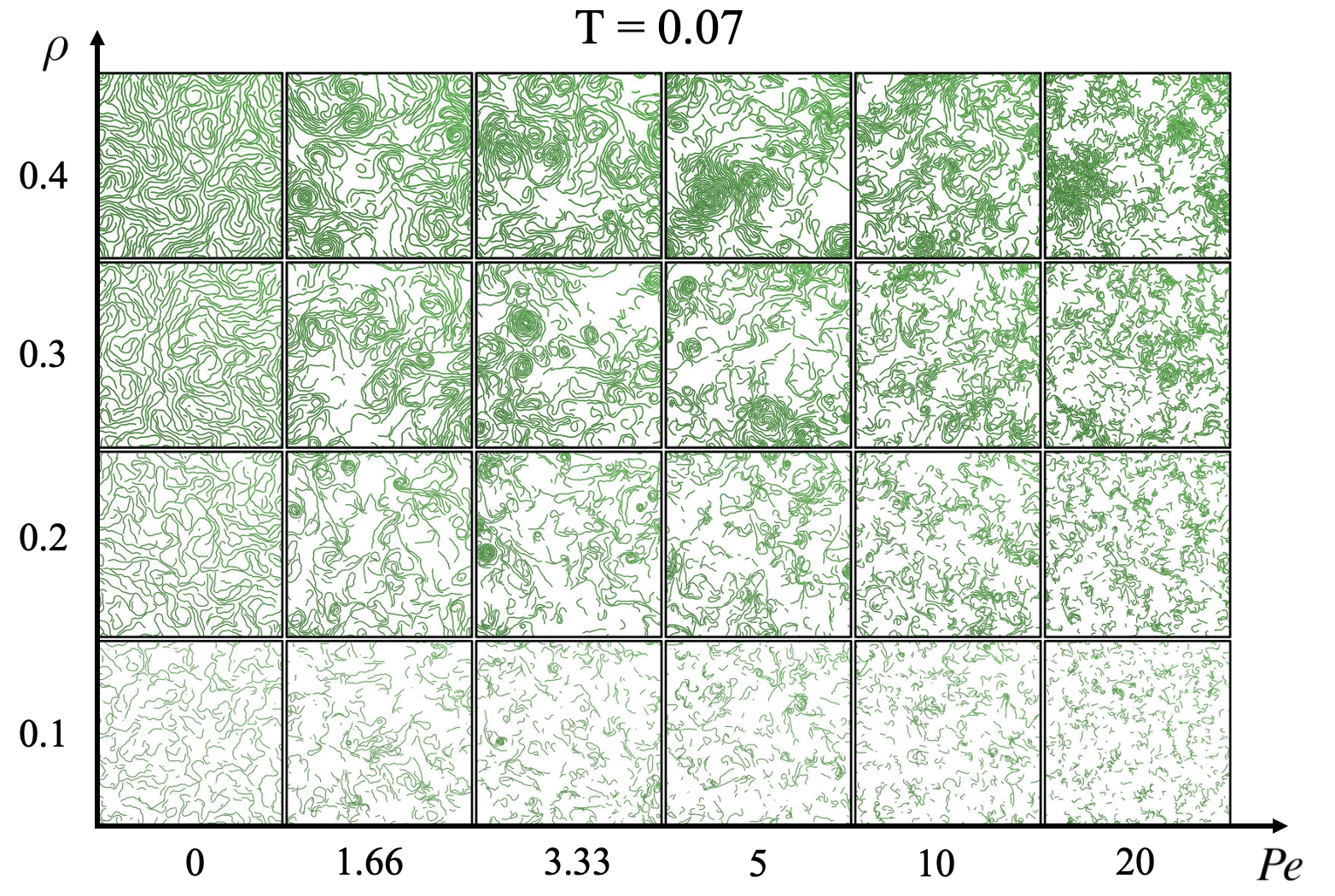}
 \includegraphics[height=5.7cm]{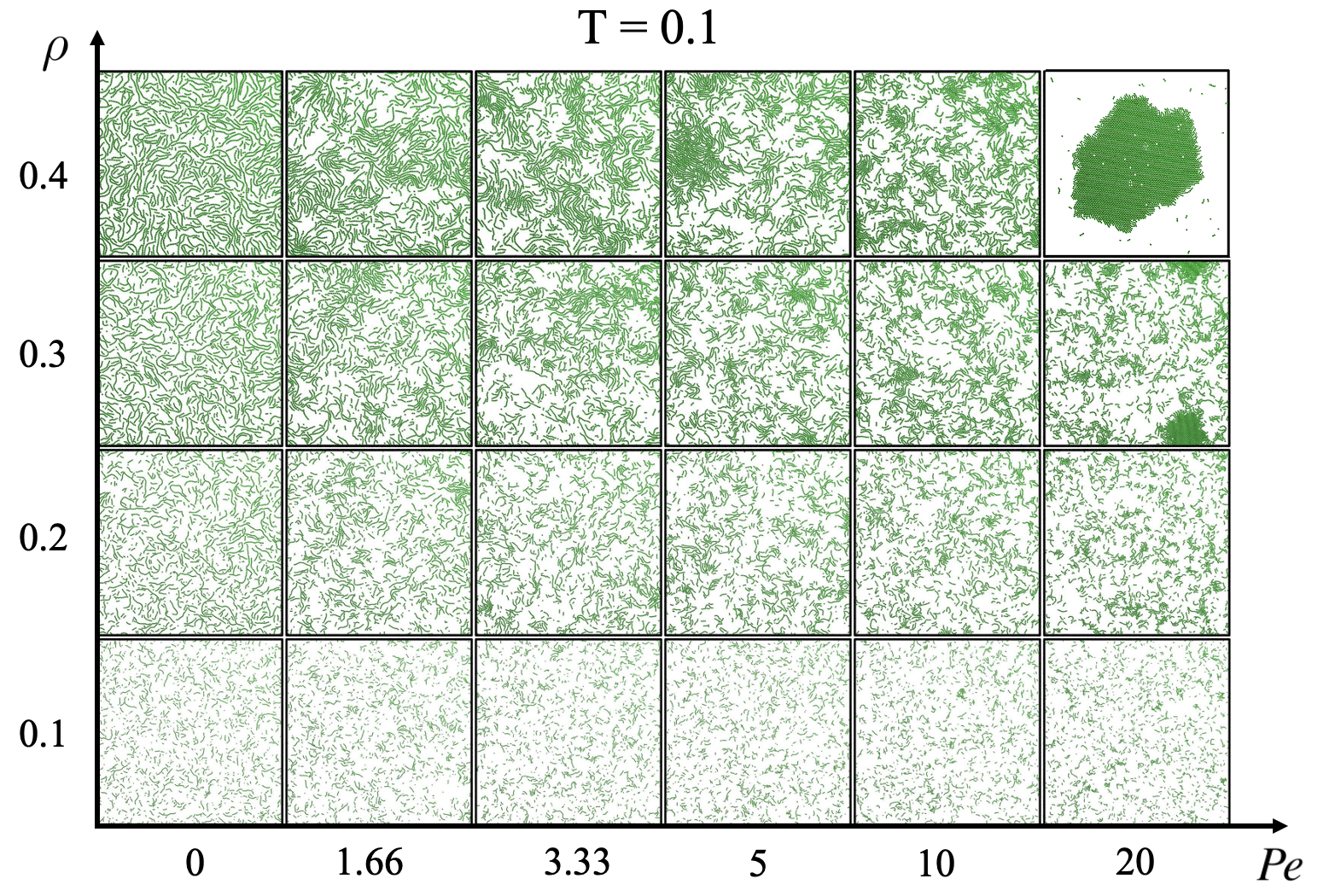}
 \caption{
 Steady state configurations, as a function of activity and density, at temperature $T=0.07$ (top panel) and $T=0.1$ (bottom panel). Activity increases horizontally (from left to right) and density increases vertically (from bottom to top).
 }
 \label{fgr:T}
\end{figure}

\textcolor{black}{We define the steady state as the state where the total number of bonds 
is stationary in time (the same applies for the total potential energy, see Supplementary Material)}.
\textcolor{black}{We cannot exclude that at longer time intervals a coarsening (or phase separation) could occur, but our evidence suggests that the system enters a stationary state where all static quantities  (such as chain length and cluster size distributions) do not change over time. }
We start by analysing the phase behaviour of the suspension when varying activity and density. Fig. \ref{fgr:T} reports two panels, each showing snapshots taken once the system was in steady state: the top one represents the system at a lower temperature (T = 0.07) and the bottom one at a higher temperature (T = 0.1).

Passive particles (left-most column) self-assemble into uniformly space-distributed linear chains (independently of temperature and density). 
Increasing activity, active particles self-assemble into linear chains that aggregate with each other. These dense and compact structures are more clearly visible at the highest density (top row in both panels of Fig. \ref{fgr:T}). At the lowest temperature, aggregates appear even at lower densities (bottom rows of top panel of Fig. \ref{fgr:T}). For the largest simulated activity  $Pe=20$ (at the highest temperature, bottom panel), the system forms a compact and ordered structure.

\textcolor{black}{ 
Counterintuitively, this compact and ordered structure forms at the highest temperature:  this is coherent with the fact that shorter chains are present at this temperature. This is further supported by the observation of this structure at the highest activity, when chains are the shortest.
}

\textcolor{black}{We underline that the spinning crystalline phase differs from the well known MIPS phase for many reasons. First, this compact structure is rotating and translating, while the dense phase in MIPS does not rotate. Second, the spinning crystalline phase is composed of chains of different lengths instead of single particles as in the MIPS phase. Third, the spinning crystalline phase is characterized by a monocrystalline structure instead of a polycrystalline one, as in the MIPS case.}
\textcolor{black}{To better characterize the region of the state diagram where the crystal is observed, we report a  zoom of the state diagram  in the Supplementary Material.}

At the lowest temperature and highest densities (rows $\rho=0.4$ or $\rho=0.3$ of top panel of Fig. \ref{fgr:T}), especially when activity assumes low or mid-range values (such as $Pe=1.66,\ 3.33,$ or $5$), chains aggregate forming spirals which are rotating at a finite angular velocity, \textcolor{black}{reminiscent of the recently experimentally observed spirals in driven actin filaments on a motility assay ~\cite{schaller2010}. We note that these spirals differ on the basis of their structures from the vortices detected in experiments of active filaments in ~\cite{Sciortino2021pattern}.}
These aggregates are very different from the density fluctuations observed in the suspensions of purely repulsive active Brownian particles (MIPS)\cite{Stenhammar2015}. For this reason, we define this novel state with the acronym MISP, i.e., Motility-Induced SPirals. 

\textcolor{black}{The most common aggregated states are temporary chains of different length or a compact spinning crystal. In the former case, i.e. when particles form a chain, they are able to adjust their propulsion direction while maintaining the bond with neighboring particles. However, if a particle's propulsion direction changes such that the particle can no longer stay connected to its neighbor, the chain breaks. This constrains how much particles in a chain can change their propulsion direction. On the other hand, a crystal is a more compact structure, even though it consists of several chains merged together. In this case, if particles change their propulsion direction enough to break the bonds within the chain they belong to, they are trapped by their neighbors and the compact structure does not break (unless they are located at the outer surface of the crystal and are free to move away).
}

\subsection{Chain and active spiral phase}

\begin{figure}[h!]
 \centering
 \includegraphics[height=7.cm]{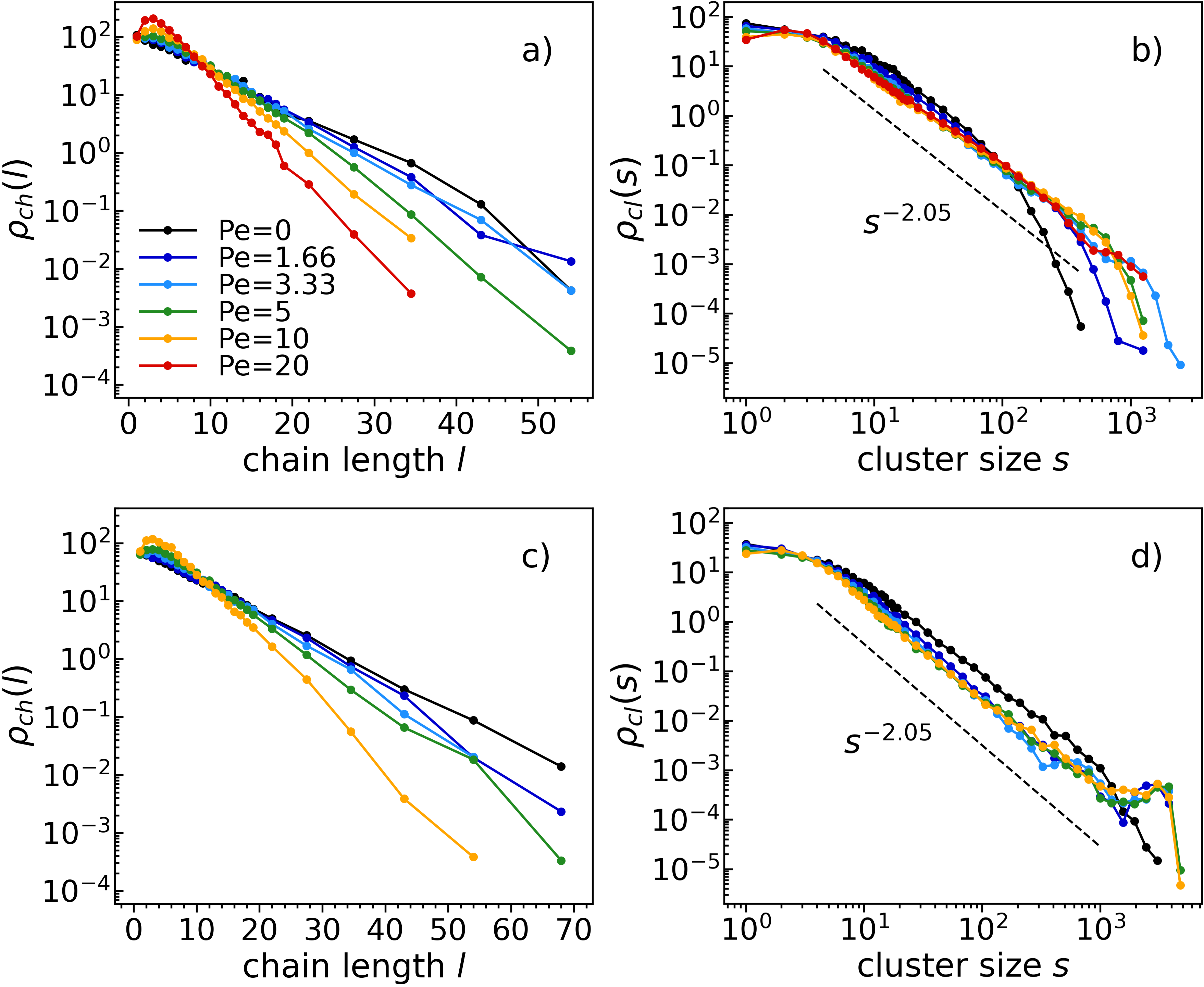}
 \caption{
 \textcolor{black}{Chain length distributions (left) and cluster size distributions (right) at temperature $T=0.1$, density $\rho=0.3$ (top) and density $\rho=0.4$ (bottom), and all Péclet numbers studied (as indicated in the legend). Note that bottom panels do not include the case $Pe = 20$ (red line), where the system is in a crystalline phase}.
 }
 \label{fgr:CLD}
\end{figure}

In either passive or active systems, particles self-assemble into linear chains whose length distribution decays exponentially \textcolor{black}{(see Fig. ~\ref{fgr:CLD} a,c)}. In active systems, chains tend to be shorter than in the corresponding passive system. In particular, the higher the activity, the shorter the chains. \textcolor{black}{This is observed at all temperatures and densities within the studied ranges (see Supplementary Material for the chain length distributions and cluster size distributions at different temperatures and densities).} Therefore, activity clearly affects chain formation. 

Moreover, as activity increases, a consistently more pronounced peak is observed at small chain lengths, likely due to the higher diffusivity of small chains as compared to long ones. \textcolor{black}{This faster diffusion of smaller chains likely results in their quicker assembly with other particles, leading to a smaller number of small chains than what would be expected from the exponential trend.
}

On top of that, as density increases, chains aggregate forming clusters. At the highest density (see Fig. ~\ref{fgr:CLD} d), both passive and active clusters exhibit percolation. In fact, all cluster size distributions follow the same power law: $\rho_{ch}(s)\sim s^{-\tau}$ with $\tau=2.05$, which is consistent within the random percolation universality class ~\cite{stauffer1992}. 

\textcolor{black}{
At the highest density (Fig. ~\ref{fgr:CLD}d), the cluster size distributions for all active systems present a peak for large cluster sizes (of the order of the total number of particles). A similar behavior has already been observed by the authors of Ref. ~\cite{Fily2013}, which have demonstrated that the cluster size distribution of a system of self-propelled soft disks exhibits a peak when the system phase separates. The peak corresponds to a cluster size equal to the average number of particles in the dense phase. On the contrary, the dilute phase contributes to the same cut-off power law observed in the homogeneous state.
\\
Using a kinetic model in a finite- size system of active particles, the authors of Ref. ~\cite{Peruani2006,Peruani2013} have quantitatively demonstrated that active systems can exhibit not only an individual phase (characterized by  a cluster-size distribution dominated by an exponential form) but also a clustering phase, characterized by a non-monotonic cluster-size distribution. In the latter case, a peak appears towards the tail of the distribution, signature of particles aggregating in one large cluster. 
We observe the same features in our dense active system.}

\textcolor{black}{Instead, at a lower density (see Fig. ~\ref{fgr:CLD} b), only the clusters in the more active systems percolate. Indeed, the passive cluster size distribution follows an exponential power law, indicating that clusters simply coincide with chains.}

\textcolor{black}{
As clearly reported in Fig. ~\ref{fgr:CLD}, neither the chain length nor the cluster size distribution show a relevant density-dependent behavior. For this reason, from now onward, we will mostly present our results at density $\rho=0.4$ (indicating when not otherwise).}

\begin{figure}[h!]
 \centering
 \includegraphics[height=4.5cm]{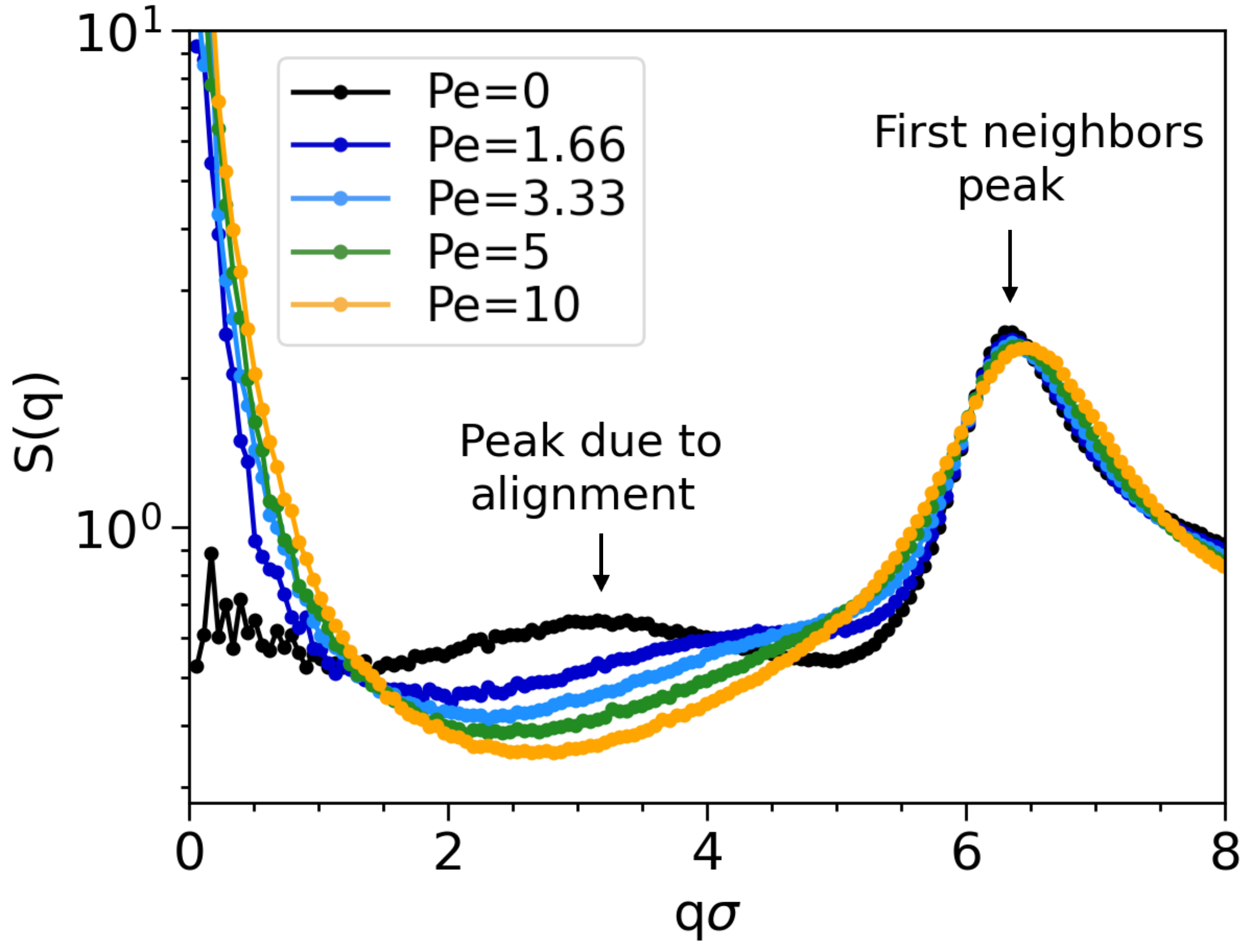}
 \caption{
 Structure factors $S(q)$ at temperature $T=0.1$, density $\rho=0.4$, and all Péclet numbers studied, except $Pe = 20$, where the system is in a crystalline phase (see legend). The peak at $q\sigma \sim3$, indicating chain alignment, disappears as activity increases. This observation holds  also at the other studied temperature and densities. 
 }
 \label{fgr:sq}
\end{figure}

\begin{figure}[h!]
 \centering
 \includegraphics[height=4.5cm]{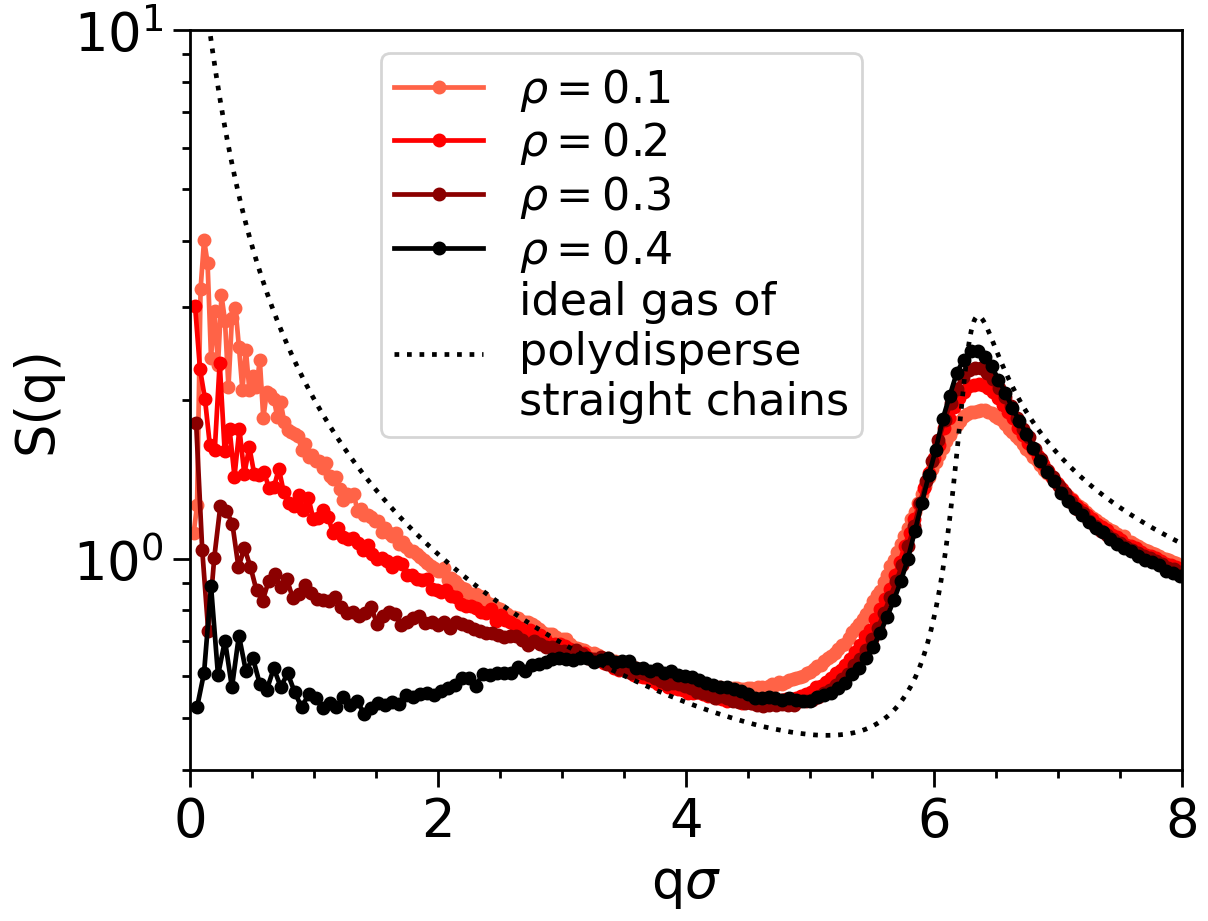}
 \caption{
 \textcolor{black}{Structure factors $S(q)$ for the passive system at temperature $T=0.1$ and varying density as indicated in the legend. The dotted black line represent the structure factor of an ideal gas of polydisperse straight chains.}
 }
 \label{sq_density}
\end{figure}

Even though clusters percolate in both passive and active systems, their structures differ significantly due to activity. Fig. ~\ref{fgr:sq} illustrates the structure factor $S(q)$ at different Péclet numbers. As expected, due to the excluded volume effects, $S(q)$ oscillates with periodicity set by the diameter (first neighbors peak in Fig. ~\ref{fgr:sq}).

An important finding in the $S(q)$ of passive systems is the presence of a peak at $q \sigma\approx3$, which indicates alignment within chains. As activity increases, the peak shifts towards that of the first neighbors and, so, disappears. Its disappearance implies that chains are aggregating among each other instead of being uniformly distributed, as a characteristic distance between chains is no longer evident. 

\textcolor{black}{
The peak at $q\sigma\approx3$ also vanishes when decreasing the density, implying that chain alignment cannot generate a sufficiently strong signal when the system is too diluted. Fig. ~\ref{sq_density} compares the $S(q)$ of passive systems at different densities with the $S(q)$ of an ideal gas of polydisperse straight chains (dotted black line). In the last case, where chains are straight and non-interacting, the peak at $q\sigma\approx3$ is not expected.
} 

\textcolor{black}{Chaining manifests in the non-negligible values of $S(q)$ at small $q$. In particular, higher values are observed in the active case compared to the passive one (see Fig. ~\ref{fgr:sq})). In the passive case, higher values are observed at lower densities (see Fig. ~\ref{sq_density}), with particularly elevated values in the case of the ideal gas of polydisperse straight chains.}

\textcolor{black}{Non-negligible values of $S(q)$ at small $q$ are present. In particular, higher values are observed in the active cases (see Fig.  ~\ref{fgr:sq}). In the passive case, higher values are observed at lower densities (see Fig. ~\ref{sq_density}), with particularly elevated values in the case of the ideal gas of polydisperse straight chains.}

\textcolor{black}{
When analyzing the structure factor at different Péclet numbers, we do not observe any significant signal indicating the presence of spiral structures (see Supporting Material for the analysis of the structure factor of spiral configurations).
}

\begin{figure}[h!]
 \centering
 \includegraphics[width=0.46\textwidth]{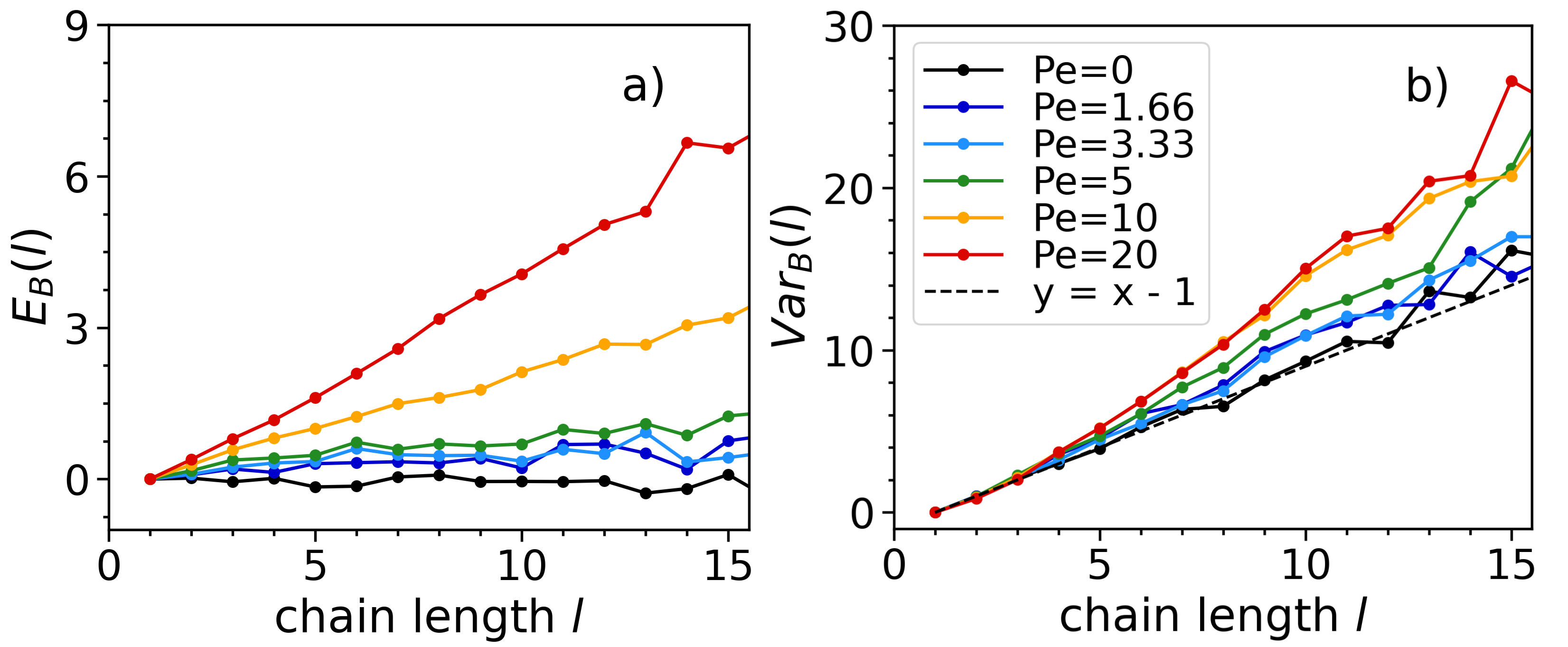}
 \caption{
 $E_B(l)$ (left) and $Var_B(l)$ (right) at temperature $T=0.1$, density $\rho=0.3$, and all Péclet numbers studied (see legend).
 As activity increases, $E_B(l)$ and $Var_B(l)$ take larger value consistently. This observation holds also for the other studied temperature and densities.
 }
 \label{fgr:PropulsionVectors}
\end{figure}

To better understand chaining, we study the dynamics of the bonding process, to unravel whether particles have a tendency to self-assemble into chains with similar or opposite orientation and whether such tendency is related to activity.
\textcolor{black}{
We will present our results at density $\rho=0.3$ at which the system (for the chosen temperature and activity range) is never in a crystalline state. This allows us to present the full range of simulated activities. 
}
At low activities, $E_B(l) \sim0$ and $Var_B(l) \sim l-1$ (see Fig. \ref{fgr:PropulsionVectors}). This means that  $B_i$ follows a Bernoulli distribution with equal probability of success (particle placed with the same orientation of the preceding one) and failure (particle placed with the opposite orientation of the preceding one). Hence, attraction dominates over activity in the bonding process, leading self-assembly to occur randomly regardless of the particle orientations. 

As activity increases, Fig. \ref{fgr:PropulsionVectors} shows that both $E_B(l)$ and $Var_B(l)$ consistently take larger values. This indicates, for every bonding event, an increase of the probability of two particles to self-assemble with the same orientation and a decrease of the probability to self-assemble with opposite ones. In this instance, self-assembly occurs favouring an alignment of the propulsion vectors. \textcolor{black}{Hence, activity affects bonding, as we observe the propulsion vectors of two particles aligning when forming a bond.}

\begin{figure}[h!]
\centering
\includegraphics[height=4.35cm]{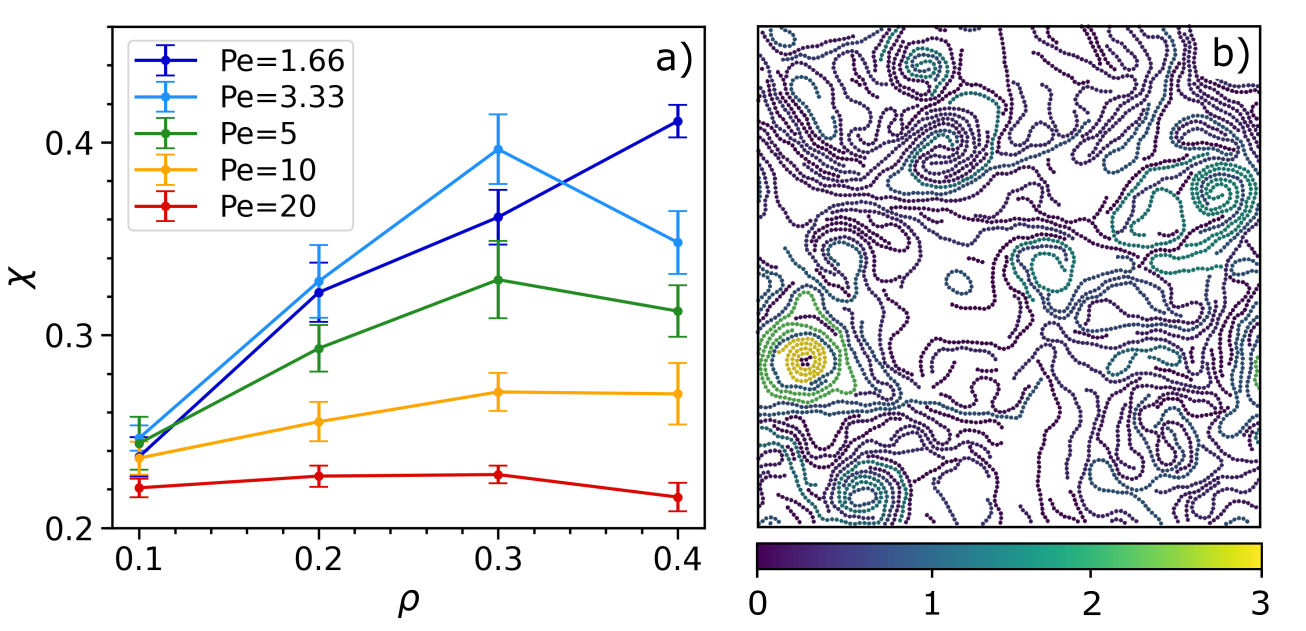}
\caption{
\textcolor{black}{a) Average turning number $\chi$ as a function of density $\rho$ at temperature $T=0.07$ and all P\'eclet numbers studied at this temperature. b) Typical spiral configuration ($T=0.07$, $Pe=1.66$, and $\rho=0.4$) with particles color-coded according to the absolute value of the spiral number $|\chi_i|$.} 
}
\label{turning number}
\end{figure}

At the lowest temperature and highest values of density and activity, chains aggregate forming rotating spirals. A movie representing spiral formation is shown in the Supporting material. The pathway for spiral formation is characterised by few steps: 1) ABBPs self-assemble to form long chains; 2) chains aggregate due to their persistent velocity; 3) due to  combined density and temperature effects, chains merge forming spirals, that spin due to alignment within the chains. 

\textcolor{black}{
We characterise the MISP state via the average turning number $\chi$ (being $\chi$ small when spirals are not present and large  when spirals are present in the system). Setting the temperature at the lowest value, where we know the system can be in a spiral state, we report the average turning number $\chi$ as a function of density $\rho$ for different P\'eclet numbers (Fig.~\ref{turning number} a).}

\textcolor{black}{
As expected, spiral formation happens at higher density and smaller activity values. This is because when the system is too diluted (low density), chains do not need to compete for space and move freely, and when activity is too high, chains are too short to coil into a spiral.}
\textcolor{black}{In the Supplementary Material we also show the probability density distribution of the average turning number.} 

\textcolor{black}{
To understand the reason why the values of the turning number reported in Figure \ref{turning number} a) are so small, we plot a typical snapshot of the system in a spiral state
in Fig. \ref{turning number} b):   particles color-coded according to the absolute value of the spiral number.  Spirals  are usually formed by more than one chain and only few of them are able to form strongly wound-up conformations. The reason is that these structures are quite unstable (unfolding and folding all the time) while also varying in size.
}

\textcolor{black}{As in Ref.~\cite{turning_numer1}, we have also computed the absolute value of the spiral number, averaging over all chains and over all configurations (Figure 8 of the Supplementary Material). 
Comparing the  spiral number to  the turning number computed for the same system, 
we observe  they coincide, being both very small  for all the chosen parameters (as shown in the snapshot of Fig. \ref{turning number} b).}

\subsection{Spinning crystalline cluster phase}

\textcolor{black}{At the highest temperature, density and activity (top-right configuration of bottom panel of Fig. \ref{fgr:T}), the system forms a crystalline cluster.}
\textcolor{black}{ Crystalline clustering is a two-step self-assembly process (first particles self-assemble into chains, next chains self-assemble into a cluster). For a more comprehensive overview of this process, in the Supplementary Material, we show local density distribution computed for the location of the centers of mass of the chains in the crystalline configuration, together with a typical snapshot showing their location in the crystalline cluster.}

\begin{figure}[h!]
\centering
\includegraphics[height=4.cm]
{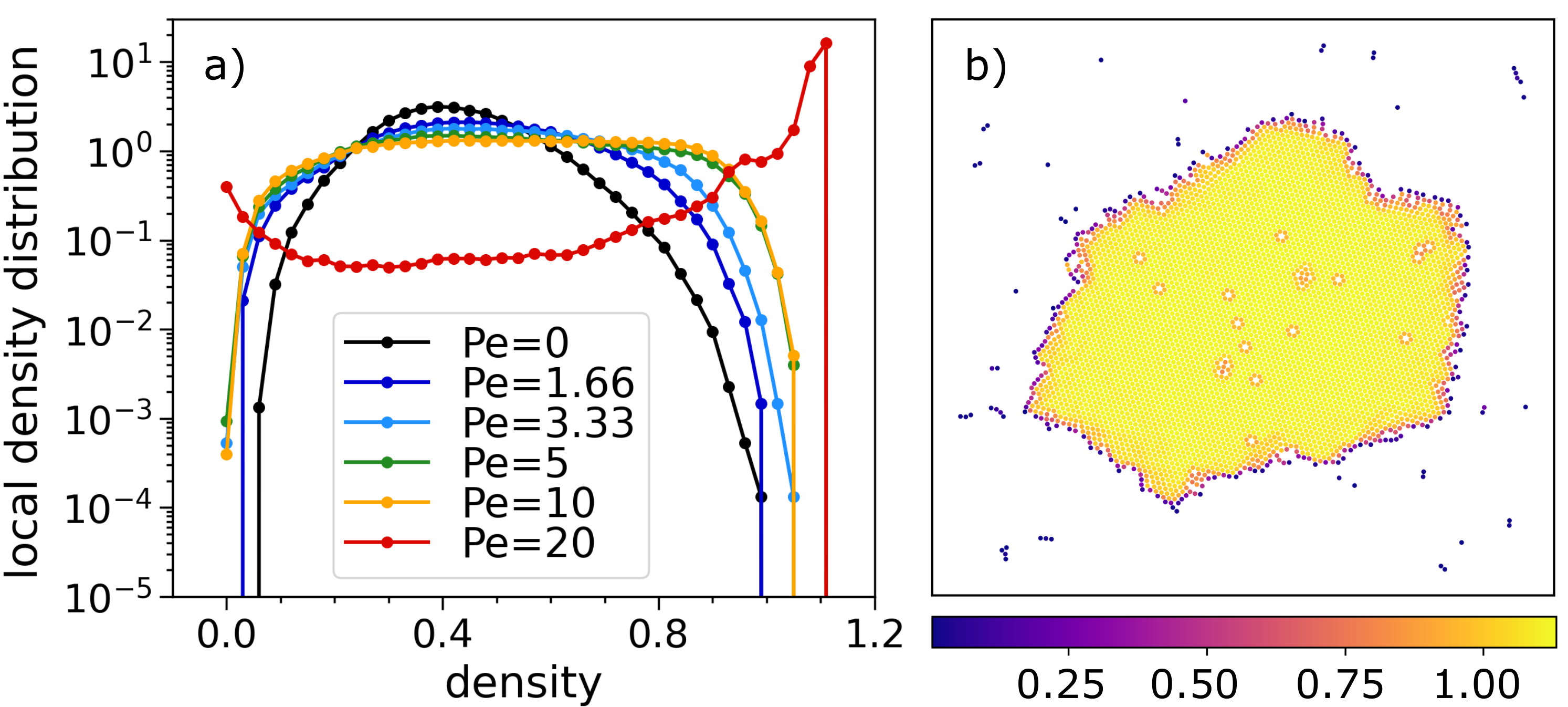}
 \caption{
 a) Local density distributions at temperature $T=0.1$ and density $\rho=0.4$. Péclet numbers vary according to the legend. At the highest Péclet number (red color), the distribution is characterized by two distinguished peaks, which indicate phase separation. \textcolor{black}{b) Crystal steady-state configuration with particles color-coded according to the reciprocal volumes of the associated Voronoi cells.}
 }
 \label{fgr:LDD}
\end{figure}

\textcolor{black}{
Fig. \ref{fgr:LDD} a) shows the local density distributions (evaluated performing a Voronoi tessellation of the system) at the highest temperature and density. A phase separation takes place at the highest activity (red line), as shown by the rise of two distinguished peaks.}
\textcolor{black}{ The bimodal behavior of the distribution is visible even though the distribution assumes non-zero values at mid-range densities, which is due to the contribution of particles located at the boundary of the crystalline cluster. Fig. \ref{fgr:LDD} b) shows all particles of the crystalline configuration color-coded according to the reciprocal volumes of the associated Voronoi cells, clearly illustrating their influence on the local density distribution.
}

\begin{figure}[h!]
 \centering
 \includegraphics[height=8.8cm]{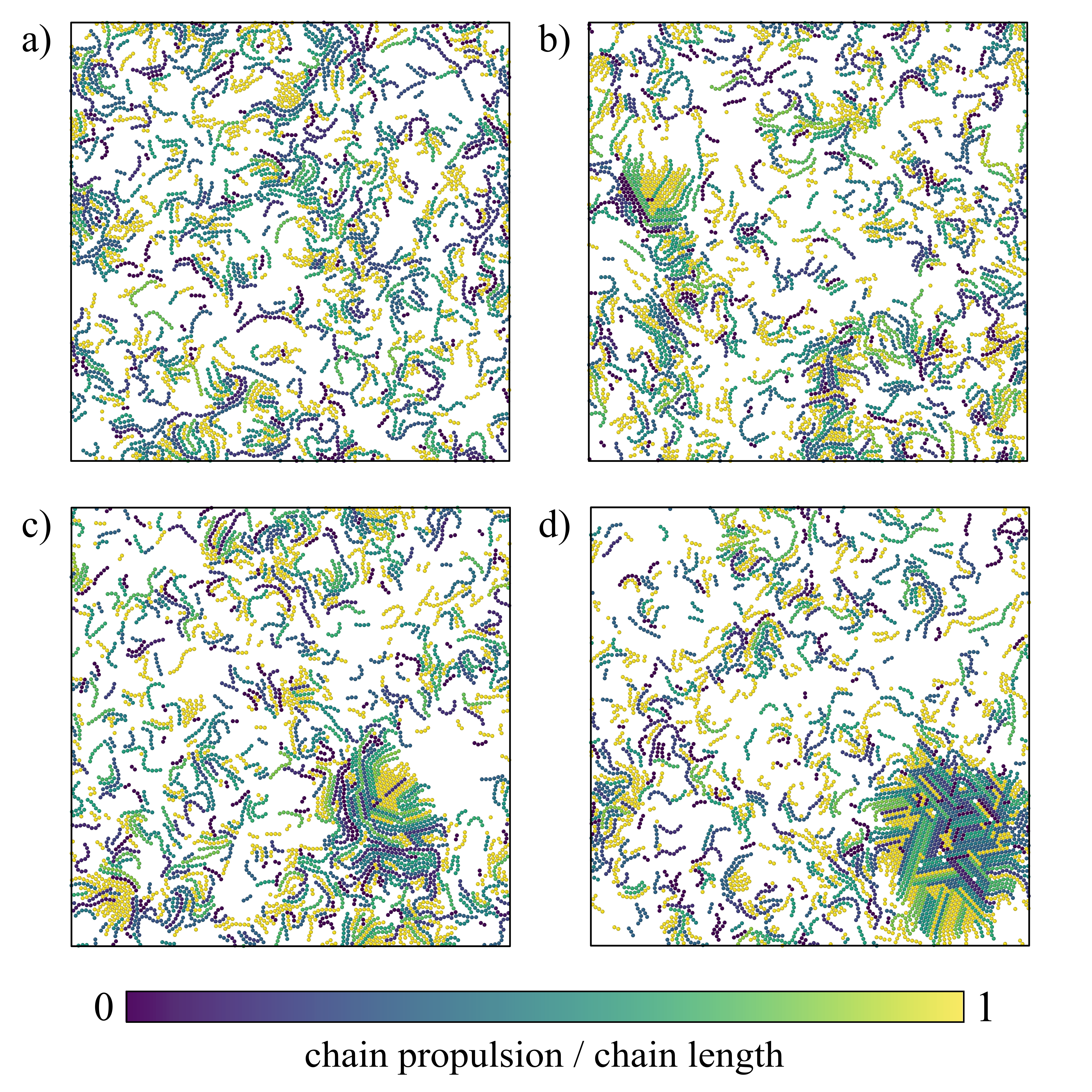}
 \caption{Snapshots taken along the crystallization process. Particles belonging to the same chain are depicted with the same color. A different color indicates a different value of the chain propulsion (as introduced in section \ref{sims}) divided by the chain length. a) Chains are forming and aggregating but there is not a stable nucleus. b) A stable nucleus is formed with a head (bluish chains) and a tail (yellowish chains). c) The nucleus is moving and aggregating chains in the head. d) The growing nucleus becomes a stable crystalline structure.}
\label{fgr:cristallo_formation}
\end{figure}

Interestingly, the pathway towards crystal formation follows the steps reported in Fig. \ref{fgr:cristallo_formation} (also a movie of the entire process is shown in Supporting material). 

Fig. \ref{fgr:cristallo_formation} a) shows an initial state where chains start to form and aggregate but not in a stable way. Fig. \ref{fgr:cristallo_formation} b) shows a stable cluster of chains with a head (bluish chains) and a tail (yellowish chains). Yellowish chains are chains where particles are pointing all in one direction and so have a non  zero chain propulsion.
Fig. \ref{fgr:cristallo_formation} c) shows that the pushing chains in the tail allow the cluster to explore the system, leading to its growth due to the aggregation of other rather slow chains.
All chains aggregate in a compact way and activity helps to anneal defects present in the cluster increasing its crystalline order (Fig. \ref{fgr:cristallo_formation} d).

\begin{figure*}[h!]
 \centering
 \includegraphics[height=6.2cm]{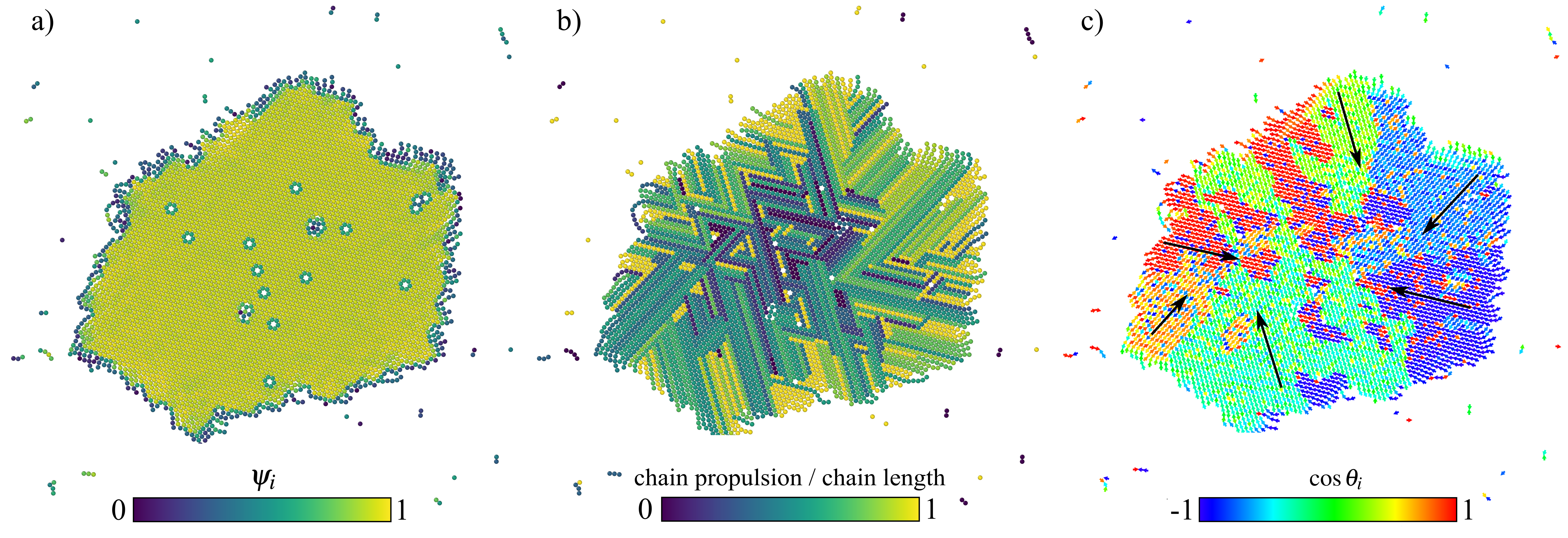}
 \caption{Crystalline structure formed at the highest temperature ($T = 0.1$), density ($\rho = 0.4$), and Péclet number ($Pe = 20$). a) Particles are colored according to the value of the crystalline order parameter $\psi_i$. While $\psi_i=1$ represents perfect order, $\psi_i=0$ no order at all. b) Particles belonging to the same chain are depicted with the same color. A different color indicates a different value of the chain propulsion (as introduced in section \ref{sims}) divided by the chain length. While $1$ indicates all particles are arranged in the chain with same orientation, $0$ with orientations alternated.
 c) Particles are colored according to the value of the orientation vector along the $x$-axis. Note that $\cos\theta_i=1$ indicates self-propulsion toward the right side of the box, $\cos\theta_i=-1$ toward the left side, and $\cos\theta_i=0$ toward the top or bottom side. The big black arrows indicate the main direction of self-propulsion for each domain. 
 }
\label{fgr:crystal}
\end{figure*}

Once the crystal is formed, it is interesting to notice that it translates and rotates.
To characterise its structure, we compute several properties.
Fig. \ref{fgr:crystal} depicts this steady state configuration in three different panels. 

In Fig. \ref{fgr:crystal} a), particles are colored according to the value of the crystalline order parameter $\psi_i$ (whose averaged value over all particles is $\psi\approx0.87$). In the core of the dense structure, particles are arranged as in a hexagonal lattice with disclinations. All particles not belonging to the crystal are  monomers, dimers or trimers. 
In Fig. \ref{fgr:crystal}b), particles are colored with the same color when belonging to the same chain and according to the value of the chain propulsion (as introduced in section \ref{sims}) divided by the chain length. In the core of the dense structure, particles are arranged in straight chains, with alternated orientations  in the innermost region and similar orientation in the outermost regions. This is due to the fact that, once the crystal core is formed, particles aggregate to it at the interface (see crystallization process in Fig. \ref{fgr:cristallo_formation}). In Fig. \ref{fgr:crystal} c), particles are colored according to the value of the orientation vector $\eta$ along the $x$-axis. We note the presence of domains where particles are self-propelling in the same direction. The direction of such domains are indicated with black arrows. This results in an applied torque to the crystalline cluster. Thus, the crystalline cluster has a finite angular velocity other than a translating motion of its center of mass dictated by the evaporating front.

\section{Conclusion}

We investigate the phase behaviour of a model system made of active Brownian particles with two opposite-located short-range attractive sites.
Our work explores the role of activity, temperature and density in the process of polymerization of active patchy particles in linear chains.

\textcolor{black}{If particles are active, they self-assemble into chains which then aggregate} (as opposed to uniformly space-distributed chains observed in the passive corresponding systems), forming from motility-induced spirals (lowest temperature and higher densities) to crystalline clusters (highest temperature, density and activity). 

To characterise the structural features of the aggregated chains, we evaluate the cluster size distributions on an energetic and geometric basis. In particular, the first method (energetic bonds) allows us to characterize the length of the chains in function of the density, temperature and, most importantly, activity. The presence of activity reduces the average chain length at every temperature and density combination. On the other hand, the second method (geometric bonds) is evaluated with the purpose to characterize the spatial chain aggregates. As a result, we observe the onset of a percolation phenomenon in a wide range of densities.

Then, we keep trying characterization the chain aggregates. This time we exploit a well-known quantity in the description of a system's structural properties, which is the structure factor. In all passive systems at the highest density, we observe the presence of an anomalous peak in the function that we attributed to the alignment of the chains. Furthermore, as activity increases, we observe that such peak shifts towards the first neighbors peak. This can be explained by the fact that when activity is introduced in the system, chains do not uniformly distribute but aggregate, and thus we cannot identify a characteristic length anymore.

Next, the analysis proceeds by investigating the arrangement of the particles within the chains based on the direction of the propulsion vectors. Specifically, our interest focuses on understanding whether the particles were bonding with propulsion vectors in the same or opposite direction. In systems with low activity, we find the probability of bonding in the same direction to be equal to the probability of bonding in the opposite direction, i.e., the attraction being dominant over activity in the bonding process. Interestingly, as activity increases, we note the probability of bonding in the same direction increases.
This result is in agreement with our predictions of a bonding process being mainly determined by activity.

To summarize, in passive systems, clusters are made of long, aligned and slow chains, while in active systems, clusters are made of short, aggregating and fast chains.

Finally, we focus on the formation and features of the crystalline structure observed at the highest values of temperature, density and Péclet number. In particular, we discover a significant rotation of the crystal cluster. The rotation arises from an effective torque generated by the presence of domains where particles are self-propelling in the same direction. What controls the nature of the fluid-to-solid transition in this active system surely deserves further investigation.

\textcolor{black}{
In this study, we can tune the flexibility of the chains by changing the angular aperture of the patch interaction. But in order to compare these results with those reported in our manuscript we would need to keep the bonding volume constant (since the dimerization constant depends on it). Unfortunately this cannot be done with the current functional form of the potential (which is square-well patchy) and would require us to switch to a different model (Kern-Frenkel \cite{kern2003}). This will be the subject of a future research.
}

In conclusion, the presented results demonstrate the rich dynamics and emergent phenomena in active bifunctional Brownian particles highlighting the potential for a deeper understanding of out-of-equilibrium systems, and for novel applications in colloidal science.

\section*{Supporting information}
\textcolor{black}{
We report two movies to describe the novel active states. Movie 1 represents the spiral formation for the system at $Pe=1.66$, $\rho=0.4$ and $T=0.07$. The spiral formation is characterized by a few steps: self-assembling of ABBPs in chains, chain aggregation  and formation and breakage of the spirals. Movie 2 represents the formation of the spinning cluster for the system at $Pe=20$, $\rho=0.4$ and $T=0.1$. The crystal formation is characterized by a few steps. After an initial chain formation and aggregation in an unstable way, it is possible to observe the formation of a stable cluster of chains characterized by a head (bluish chains) and a tail (yellowish chains). This comet-like cluster grows in size and order through the incorporation of new chains, while moving in the space available in the system.
}

\section*{Author Contributions}
Please refer to our general \href{https://www.rsc.org/journals-books-databases/journal-authors-reviewers/author-responsibilities/}{author guidelines} for more information about authorship.

\section*{Conflicts of interest}
There are no conflicts to declare.

\section*{Data Availability Statement}
The data that support the findings of this study are available from the corresponding authors upon reasonable request.

\section*{Acknowledgements}
C.V. acknowledges fundings IHRC22/00002 and PID2022-140407NB-C21 from MINECO. J.R. and F.S. acknowledge support by ICSC -- Centro Nazionale di Ricerca in High Performance Computing, Big Data and Quantum Computing, funded by European Union -- NextGenerationEU. The authors thank José Martín-Roca, Horacio Serna, Giulia Janzen and Daniel A. Matoz-Fernandez for insightful suggestions.

\bibliography{rsc}
\bibliographystyle{rsc}

\end{document}